\documentclass[12pt]{ISAS_IR}

\pdfoutput=1

\usepackage[english]{babel}
\usepackage[latin1]{inputenc}
\usepackage{ifpdf}
\usepackage{pgf}
\usepackage{tikz}
\usepackage{calc}
\usepackage{amsmath}
\usepackage{amssymb}
\usepackage{stfloats} 
\usepackage{flushend}
\usepackage{calc}
\usepackage{tikz}
\usetikzlibrary{arrows,decorations}
\usepackage{microtype}
\usepackage{algorithm2e}
\usepackage{algpseudocode}
\usepackage{trfsigns}
\usepackage{amsmath}
\usepackage{amssymb}
\usepackage{epsfig}
\usepackage{graphicx}
\usepackage{psfrag}
\usepackage{verbatim}
\usepackage{wrapfig}
\usepackage{multirow}
\usepackage{microtype}
\usepackage{lmodern}

\date{}

\def\vec#1{\underline{#1}}

\def\mat#1{{\mathbf #1}}

\def\1_2{{\frac{1}{2}}}

\def\ODE{system of ordinary first-order differential equation}

\def\SNDS{stochastic nonlinear dynamic system}
\def\CoDiCo{coupled discrete-continuous}

\def\lhs{left-hand-side}
\def\rhs{right-hand-side}

\def\etavec{\vec{\eta}}
\def\ParVec#1{\etavec_c(#1)}

\def\ParVecT#1{\etavec_c^T(#1)}
\def\ParVecDot#1{\dot{\etavec}_c(#1)}

\def\pvecT{\vec{p}^T(\vec{x},\gamma)}

\def\Pmat{\mat{P}(\gamma)}
\def\bvec{\vec{b}(\gamma)}
\def\Qmat{\mat{Q}(\gamma)}

\def\NewR{{\rm I\hspace{-.17em}R}}

\def\Eq#1{(\ref{#1})}

\def\Fig#1{Figure~\ref{#1}}

\def\Box#1{{\fboxsep2pt\fbox{$#1$}}}

\usepackage{theorem}
\theoremheaderfont{\bfseries}
\theoremstyle{plain}
{\theorembodyfont{\itshape} }
{\theorembodyfont{\itshape} }
{\theorembodyfont{\itshape\small} }
{\theorembodyfont{\sffamily}}
{\theorembodyfont{\itshape} \newtheorem{Remark}       {Remark}    [section]}
{\theorembodyfont{\itshape} }

{\noindent{\scshape Proof.}}
{\hspace*{\fill}$\square$}

\newlength\EqLen

\def\ScaleInner#1{%
\settowidth{\EqLen}{#1}
\ifdim\EqLen < \columnwidth%
\begin{equation*}%
\begin{minipage}{\EqLen}#1\end{minipage}%
\end{equation*}%
\else%
\begin{equation*}%
\resizebox{0.99\columnwidth}{!}{\begin{minipage}{\EqLen}#1\end{minipage}}%
\end{equation*}%
\fi%
}%

\def\Scale#1
  {
  \ScaleInner{$ #1 $} 
  }

\begin{document}

\begin{frontmatter}

\title{\LARGE\bfseries\sffamily Progressive Gaussian Filtering}

\author{%
Uwe~D.~Hanebeck, Jannik~Steinbring\\
Intelligent Sensor-Actuator-Systems Laboratory (ISAS)\\
Institute for Anthropomatics\\
Karlsruhe Institute of Technology (KIT), Germany\\
{\tt uwe.hanebeck@ieee.org,jannik.steinbring@student.kit.edu}
}

%
%

\begin{abstract}
In this paper, we propose a progressive Bayesian procedure, where the measurement information is continuously included into the given prior estimate (although we perform observations at discrete time steps). The key idea is to derive a system of ordinary first-order differential equations (ODE) by employing a new coupled density representation comprising a Gaussian density and its Dirac Mixture approximation. The ODE is used for continuously tracking the true non-Gaussian posterior by its best matching Gaussian approximation. The performance of the new filter is evaluated in comparison with state-of-the-art filters by means of a canonical benchmark example, the discrete-time cubic sensor problem.
\end{abstract}

\end{frontmatter}

%
%

\section{Introduction} \label{Sec_Introduction}

    %
%
We consider state estimation in discrete-time \SNDS s.
%
%
Thanks to their simplicity, Gaussian filters, i.e., filters representing all state densities by Gaussians, are an attractive tool for solving this type of estimation problem.
%
%
However, in general, their full estimation performance is not exploited due to additional assumptions and simplifications.

%
%
In this paper, we focus on Gaussian filters that operate by finding the best-matching Gaussian to the true posterior by means of an explicit shape optimization after every processing step. A filter of this type is called Gaussian-assumed density filter (GADF).

%
%
In case of the filter step, the actual measurement is considered during the approximation, which leads to better results than the commonly used type of Gaussian filter, the so called Linear Regression Kalman Filter (LRKF) \cite{lefebvre_linear_2005}. LRKFs approximate the (non-Gausian) joint density of state and measurement corresponding to the prior density, the noise density, and the given nonlinear measurement equation by a jointly Gaussian density, which is independent of the actual measurement and typically is only a very rough approximation of the true joint density.

%
%

Examples of LRKFs \cite{lefebvre_kalman_2004} are the Unscented Kalman Filter (UKF) \cite{julier_new_2000}, and its scaled version \cite{julier_scaled_2002}, its higher-order generalization \cite{tenne_higher_2003}, a generalization to an arbitrary number of deterministic samples placed along the coordinate axes
\cite{IFAC08_Huber}, 
filters performing an analytic or semi-analytic calculation of the required moments
\cite{ACC11_Huber}, 
\cite{MFI10_Beutler} 
based on a decomposition into parts that can be calculated in closed form or via a sample approximation
\cite{Fusion09_Beutler}, 
and filters based on numerical integration for calculating the required nonlinear moments of the prior Gaussian density \cite{ito_gaussian_2000}.

%
%
Of course, GADFs are more complicated to implement in comparison to the LRKFs and there are various options for minimizing the shape deviation between the true posterior and its Gaussian approximation. One option is to employ moment matching, i.e., using the mean and covariance matrix of the true posterior as parameters for the desired Gaussian, as this is known to minimize the Kullback-Leibler distance between the two densities. Unfortunately, in the case of nonlinear measurement equations and the corresponding complicated Likelihood function, it is in general not a simple task to calculate mean and covariance matrix of the true posterior, as analytic solutions are a rare exception. In most cases, numerical integration over the true posterior, i.e., the product of the (Gaussian) prior density and the Likelihood function, is in order, such as Monte Carlo integration \cite{kotecha_gaussian_2003}.

%
%
In this paper, we propose a progressive Bayesian procedure for Gaussian-assumed density filtering, where the measurement information is continuously included into the given prior estimate (although we perform observations at discrete time steps).
%
%
The first progressive filtering procedure has been developed in 2003
\cite{SPIE03_HanebeckBriechle-ProgBayes} 
for state estimation with a Gaussian Mixture representation in the scalar case, where a homotopy continuation approach was proposed for tracking the true posterior with its approximation minimizing a squared-integral distance measure. The multi-dimensional case was treated in
\cite{CDC03_Feiermann-ProgBayes}. 
A generalization of this method to various another distance measures is proposed in \cite{hagmar_optimal_2011}. Besides state estimation, the progressive processing idea has been applied to moment calculation
\cite{SPL03_Rauh} 
and Gaussian Mixture reduction
\cite{Fusion08_Huber-PGMR}. 
A homotopy-based filtering method operating on a particle representation is given in \cite{daum_nonlinear_2009}.

%
%
This paper is a further development of the procedure 
\cite{CDC03_Feiermann-ProgBayes} 
that was developed for state estimation with a Gaussian Mixture representation.
%
%
The key idea is to derive a \ODE s (ODEs) by employing a new coupled density representation comprising a continuous and a discrete part, specifically a Gaussian density and its Dirac Mixture approximation
\cite{CDC09_HanebeckHuber}. 
%
%
Dirac Mixture approximations for nonlinear estimation and filtering have been proposed for the case of scalar continuous densities in
\cite{CDC06_Schrempf-DiracMixt}, 
\cite{MFI06_Schrempf-CramerMises} 
An algorithm for sequentially increasing the number of components is given in  
\cite{CDC07_HanebeckSchrempf} 
and applied to recursive nonlinear prediction in
\cite{ACC07_Schrempf-DiracMixt}. 
Multi-dimensional Gaussian densities are treated in \cite{CDC09_HanebeckHuber}.
%
%
Of course, more complicated continuous densities can be handled with this approach such as Gaussian Mixture densities, which is outside the scope of this paper.
%
%
The \ODE s is used for continuously tracking the true non-Gaussian posterior by its best matching Gaussian approximation.

%
%

In contrast to 
\cite{CDC03_Feiermann-ProgBayes}, 
where a squared-integral distance measure between the true and the approximate posterior was minimized, here we derive ODEs for directly tracking the mean and covariance of the true posterior density by its Gaussian approximation.

%
%
The new progressive estimation method allows for arbitrary noise structures, even for noise structures that cannot easily be treated by LRKFs such as multiplicative noise.
%
%
The required integrals are solved by employing the discrete part, i.e., the Dirac Mixture, of the \CoDiCo-approximation, which allows to automatically place Dirac components solely in the interesting regions of the state space, that is the support of the true posterior.
%
%
As a result, the new filtering method is fast, efficient, and robust.
%
%
Its performance is evaluated in comparison with state-of-the-art filters by means of a canonical benchmark example, the discrete-time cubic sensor problem.

%
%

\section{Problem Formulation} \label{Sec_Prob_Form}

	%
%

We consider the general problem of estimating the hidden state of a discrete-time \SNDS{} based on noisy measurements, which consists of a prediction step (or time update) employing a system model for propagating the estimated state from time step to time step and filter step (or measurement update) for including observations taken at a given time step into the state estimate. Here, the focus is on the filter step that is typically considered harder compared to the prediction step. The insights obtained for the filter step can, however, be used for the prediction step as well.

%
%


%
%

A generative measurement equation
\begin{equation}
\hat{\vec{y}} = \vec{h}(\vec{x},\vec{v})
\label{Eq_Meas_Model}
\end{equation}
is investigated, with state $\vec{x}$, a specific measurement $\hat{\vec{y}}$, and measurement noise $\vec{v}$ with corresponding noise density $f_v(\vec{v})$.
%
%
The special case of additive noise
\begin{equation}
\hat{\vec{y}} = \vec{h}(\vec{x}) + \vec{v}
\label{Eq_Meas_Model_Additive}
\end{equation}
is also of interest, as it usually simplifies matters.

%
%

We assume that the generative model can somehow be converted to a probabilistic model represented by the conditional density $f(\vec{y}|\vec{x})$. For a given specific measurement $\hat{\vec{y}}$, this conditional density is the so called Likelihood function\footnote{The Likelihood function is not necessarily a valid density function. Although it is always non-negative, it does not necessarily integrate to one, i.e., it usually is not normalized or even cannot be normalized.} abbreviated as
\begin{equation*}
f_L(\vec{x}) = f(\hat{\vec{y}} | \vec{x})
\end{equation*}
%
%
For the case of additive noise, the Likelihood function is given by
\begin{equation*}
f_L(\vec{x}) = f_v(\hat{\vec{y}} -\vec{h}(\vec{x})) \enspace .
\end{equation*}
Other noise structures result in different Likelihood functions.

\subsection*{Gaussian Filters}

%
%

For a Gaussian-assumed density filter, we have a Gaussian prior density $f^p(\vec{x})$ that undergoes a Bayesian filter step according to the following multiplication with the Likelihood function
\begin{equation*}
\tilde{f}^e(\vec{x}) = c \cdot f^p(\vec{x}) \cdot f_L(\vec{x}) \enspace ,
\end{equation*}
where the resulting posterior density is denoted by $\tilde{f}^e(\vec{x})$. $c$ is a normalization constant ensuring that the posterior density integrates to one. The tilde is used to underline that this is the true density resulting from performing a single filter step.

%
%

Obviously, the true posterior density in general is not Gaussian anymore. In order to enable recursive processing without increase in computational complexity, the true posterior has to be approximated by a Gaussian density for the next processing step. 
%
%
The state-of-the-art for doing so will be reviewed in the next section.
	

\section{Density approximations} \label{Sec_CoupledApprox}

	%
%

Estimating the state of \SNDS s poses three challenges: 1.~For a given finite-parameter representation of the state density, the density type is usually not maintained across the filter or the prediction step, so that continuous re-approximations are necessary to allow for recursive processing. 2.~By doing so, the number of parameters of the selected representation typically increases without bound. 3.~The approximation of the true density by the selected representation cannot be efficiently performed or the true density cannot even be calculated at all.


\subsubsection*{Continuous Density Representations}

%
%

For representation purposes, continuous representations such as mixtures of Gaussians are convenient when it comes to presenting and using the results of the state estimation process. They smoothly cover the considered part of the state space, easily represent complicated density shapes, and require relatively few parameters for satisfactory approximation results.
%
%
However, re-approximating the true density resulting from a certain processing step by a continuous is usually hard. One example in the context of the Bayesian filter step is the re-approximation of the product of a prior Gaussian Mixture and a complicated Likelihood function by a posterior Gaussian Mixture. Another example concerning the prediction step is the propagation of a Gaussian Mixture through a nonlinear dynamic system.


\subsubsection*{Discrete Density Representations}

%
%

The difficulties with processing continuous densities led to the development of several types of filters employing discrete density representations over continuous domains. 
%
%
Using discrete samples instead of a continuous representation simplifies both processing steps significantly. Especially the prediction step becomes almost trivial as it just consists of propagating point values independently through the system equation. The filter step, on the other hand, is more challenging. At first, it seems simple to implement as it just consists of multiplying the discrete samples with the Likelihood function. However, it 1.~requires an explicit Likelihood function that is not always available and 2.~in a na\"ive implementation usually leads to a degeneration of the sample set as most samples obtain zero weights. 

%
%

Furthermore, it is usually difficult to assess the quality of a sample representation, which is related to specifying the number of samples required for a given task. Although summary statistics such as mean and covariance of the sample set can easily be calculated, more useful measures such as entropy or smoothness cannot easily be obtained.

%
%

Although the solution to the stochastic state estimation problem consists of fully deterministic densities, particle filters employing independent random samples become popular due to their simplicity. However, random samples of small size do not provide a homogeneous coverage of the underlying density. As a result, the samples usually do not adequately represent the underlying true density. In addition, the quality of the results depends on the specific realization and can only be assessed via Monte Carlo simulations and appropriate averaging.


\subsubsection*{Alternating Density Representations}

%
%

This includes sample-based Gaussian filters, such as the Unscented Kalman Filter (UKF) \cite{julier_unscented_2004} and the Gaussian filter in 
\cite{IFAC08_Huber}, 
where a prior Gaussian density is approximated by deterministic samples that undergo the processing step. The processing result is approximated by a Gaussian density again and the process is repeated.
%
%
This procedure cannot be easily generalized to more general continuous representations such as Gaussian Mixtures. This is due to the fact that approximating the discrete representation by a continuous representation is equivalent to density estimation from samples, which is a tough problem.


\subsubsection*{New \CoDiCo-approximation}

%
%

We now introduce a new class of density representations, the so called \CoDiCo-approximations, where two types of density approximations are maintained simultaneously. Instead of switching back and forth between representations convenient for the current task at hand, the desired representation is always available when needed.
%
%
In doing so, we can make use of the best features of both worlds. The continuous part is employed for maintaining the smoothness of the representation and for taking the derivatives with respect to the density parameters. The discrete part is employed for evaluating and comparing densities and for solving integrals in a quasi Monte-Carlo fashion.

%
%

The principle is to have an inner core representation, a continuous one called $f_c$, that controls an outer representation, a discrete one called $f_d$.
%
%
Typically, the parameters $\vec{\eta}_c$ of the inner representation are specified and lead to appropriate parameters $\vec{\eta}_d$ of the outer representation that is then used for, e.g., numerical integration.

%
%

The two approximations are tied in such a way that changing the shape of one representation directly influences the shape of the other one and its corresponding parameters. The degrees of freedom of the \CoDiCo-approximation are given by the number of parameters of the inner representation. The number of parameters of the outer approximation is a least equal to the number of parameters of the inner approximation, but in most cases will be larger.

%
%

As a specific inner representation $f_c(\vec{x},\etavec_c)$, we will focus on Gaussian densities, where the parameter vector $\etavec_c$ contains the elements of the mean and the covariance matrix. 
%
%
The outer representation is given by the Dirac Mixture approximation of the inner Gaussian calculated as described in
\cite{CDC09_HanebeckHuber}. 
%

%
%

Modifying the inner parameters of the new \CoDiCo-approximation based on employing the outer parameters for numerical integration is demonstrated in the next section.
	

\section{Progressive Filter Step} \label{Sec_FilterStep}

	%
%

The first step is to redefine the Likelihood function in order to achieve a continuous execution of the filter step. This progressive Likelihood is defined by
\begin{equation*}
f_L(\vec{x},\gamma) \enspace ,
\end{equation*}
where $\gamma$ is an artificial time with $\gamma \in [0,1]$. It is desired that
\begin{equation*}
f_L(\vec{x},\gamma) =
\begin{cases}
1 & \gamma=0 \\
f_L(\vec{x}) & \gamma=1
\end{cases}
\end{equation*}
holds.

%
%

Several options for defining progressive Likelihood functions exist. This includes progressively modifying the given nonlinear mapping $\vec{h}(\vec{x})$ of the underlying generative model \Eq{Eq_Meas_Model} or varying the noise variance as in \cite{CDC03_Feiermann-ProgBayes}.  
%
%
Here, we use the exponentiation of the Likelihood function as used in \cite{daum_nonlinear_2009}, \cite{hagmar_optimal_2011}. The modified Likelihood function is then given by
\begin{equation}
f_L(\vec{x},\gamma) = \left[ f_L(\vec{x}) \right]^\gamma \enspace .
\label{Eq_Progress_Exp}
\end{equation}

\begin{Remark}[Additive Noise]
%
%
For the special case of generative systems suffering from additive noise in \Eq{Eq_Meas_Model_Additive}, we obtain
\footnote{For the additive noise case, exponentiation is equivalent to progressive modification of the noise covariance matrix as proposed in \cite{CDC03_Feiermann-ProgBayes} since we have
\begin{equation*}
\gamma \, \left( \vec{y} - \vec{h}(\vec{x}) \right)^T \mat{C}_v^{-1} \left( \vec{y} - \vec{h}(\vec{x}) \right) 
= \left( \vec{y} - \vec{h}(\vec{x}) \right)^T \left(\frac{1}{\gamma} \mat{C}_v\right)^{-1} \left( \vec{y} - \vec{h}(\vec{x}) \right) \enspace .
\end{equation*}
}
\begin{equation*}
f_L(\vec{x},\gamma) = \exp\left\{ -\frac{1}{2} \, \gamma \, \left( \vec{y} - \vec{h}(\vec{x}) \right)^T \mat{C}_v^{-1} \left( \vec{y} - \vec{h}(\vec{x}) \right)  \right\} \enspace ,
\end{equation*}
where multipliers have been omitted to achieve $f_L(\vec{x},\gamma=0)=1$.
\end{Remark}

%
%

The second step is the continuous execution of the filter step, for now written with a general prior density $f^p(\vec{x})$ and a modified posterior density $f^e(\vec{x},\gamma)$ depending on the artificial time $\gamma$ introduced above
\begin{equation*}
\tilde{f}^e(\vec{x},\gamma) = f^p(\vec{x}) \cdot f_L(\vec{x},\gamma)
\end{equation*}
for $\gamma \in [0,1]$. For the final time\footnote{Of course, time here denotes the artificial time $\gamma$ introduced above.} $\gamma=1$, the modified Likelihood reaches the original Likelihood and as a result, the original posterior is attained by the modified posterior, i.e., $\tilde{f}^e(\vec{x},\gamma=1)=\tilde{f}^e(\vec{x})$. On the other extreme, for the start time $\gamma=0$, the modified posterior is desired to be identical to the prior density, i.e., we have $\tilde{f}^e(\vec{x},\gamma=0)=f^p(\vec{x})$. Hence, we have
\begin{equation*}
\tilde{f}^e(\vec{x},\gamma) = \tilde{f}^e(\vec{x},\gamma=0) \cdot f_L(\vec{x},\gamma)
\end{equation*}
for $\gamma \in [0,1]$.

%
%

Plugging in the continuous part of the \CoDiCo-approximation introduced in the previous Section gives
\begin{equation}
f_c^e(\vec{x},\ParVec{\gamma}) \approx \tilde{f}^e(\vec{x},\gamma) = f^p(\vec{x}) \cdot f_L(\vec{x},\gamma)
\label{Eq_fe_fp_fl_approx}
\end{equation}
for $\gamma \in [0,1]$, where left-hand-side and right-hand-side now become equal only for $\gamma=0$, but only approximately%
\footnote{As long as working with the infinite-dimensional functional representation $\tilde{f}^e$ of the true density, the two sides in \Eq{Eq_fe_fp_fl_approx} are equal as $\tilde{f}^e(\vec{x},\gamma)$ is capable of following changes in the \rhs{} exactly. On the other hand, for a finite-dimensional representation $f_c^e(\vec{x},\ParVec{\gamma})$, i.e., a density function depending on a finite-dimensional parameter vector $\ParVec{\gamma}$, the \lhs{} cannot necessarily exactly follow the changes of the \rhs{} as the product on the \rhs{} typically is not of the same density type.}
equal for $\gamma>0$.

%
%

In order to find the best matching Gaussian approximation $f_c^e(\vec{x},\ParVec{\gamma})$ in \Eq{Eq_fe_fp_fl_approx} we desired the first moments to be equal as
\begin{equation*}
\int_{\NewR^N} \vec{m} \, f_c^e(\vec{x},\ParVec{\gamma}) \, d \vec{x}
= \int_{\NewR^N} \vec{m} \, f^p(\vec{x}) \cdot f_L(\vec{x},\gamma) \, d \vec{x} \enspace ,
\end{equation*}
with $\vec{m}=[1,x,x^2]^T$. Taking the derivative with respect to $\gamma$ on both sides gives
\begin{equation*}
\int_{\NewR^N} \vec{m} \, \frac{\partial f_c^e(\vec{x},\ParVec{\gamma})}{\partial \gamma} \, d \vec{x}
= \int_{\NewR^N} \vec{m} \, f^p(\vec{x}) \cdot
\underbrace{ \frac{\partial f_L(\vec{x},\gamma)}{\partial \gamma}  }_{\dot{f}_L(\vec{x},\gamma)}  \, d \vec{x} \enspace ,
\end{equation*}
with
\begin{equation*}
\frac{\partial f_c^e(\vec{x},\ParVec{\gamma})}{\partial \gamma} 
= \underbrace{\frac{\partial f_c^e(\vec{x},\ParVec{\gamma})}{\partial \, \ParVecT{\gamma}}}_{\pvecT}
\cdot \underbrace{{\frac{\partial \, \ParVec{\gamma}}{\partial \gamma}}}_{\ParVecDot{\gamma}} \enspace .
\end{equation*}
We obtain
\begin{equation*}
\underbrace{\int_{\NewR^N} \vec{m} \, \pvecT \, d \vec{x}}_{\Pmat} \, \ParVecDot{\gamma}
= \int_{\NewR^N} \vec{m} \, f^p(\vec{x}) \cdot \dot{f}_L(\vec{x},\gamma)  \, d \vec{x} \enspace ,
\end{equation*}
which leads to 
\begin{equation}
\Pmat \; \ParVecDot{\gamma} = \bvec \enspace .
\label{Eq_Pmat_eta_bvec}
\end{equation}
%

%
%
For the specific progression from \Eq{Eq_Progress_Exp}, we obtain
\begin{equation*}
\dot{f}_L(\vec{x},\gamma) 
= \frac{\partial f_L(\vec{x},\gamma)}{\partial \, \gamma}
= \frac{\partial f_L^{\gamma}(\vec{x})}{\partial \, \gamma}
= f_L^{\gamma}(\vec{x}) \cdot \log\left( f_L(\vec{x}) \right)
= f_L(\vec{x},\gamma) \cdot \log\left( f_L(\vec{x}) \right) \enspace .
\end{equation*}

%
%

\begin{Remark}[Additive Noise]
A further simplification is achieved by focusing on the case of additive noise. Taking the derivative of the Likelihood now gives
\begin{equation*}
\dot{f}_L(\vec{x},\gamma) =
-\frac{1}{2} \, f_L(\vec{x},\gamma) \left( \vec{y} - \vec{h}(\vec{x}) \right)^T \mat{C}_v^{-1} \left( \vec{y} - \vec{h}(\vec{x}) \right) \enspace .
\end{equation*}
\end{Remark}


\section{Scalar Gaussian-Assumed Density Filter} \label{Sec_GaussFilter}

	%
%

This section is devoted to deriving specific formulas for scalar Gaussian densities.

\subsection*{Closed-form Expressions for Matrix $\Pmat$}

We start with a scalar Gaussian density for representing $f_c^e$ given by
\begin{equation*}
f_c^e(x,\ParVec{\gamma}) = w(\gamma) \, \frac{1}{\sqrt{2 \pi} \, \sigma(\gamma)} \exp\left( -\frac{1}{2} \frac{(x-m(\gamma))^2}{\left( \sigma(\gamma) \right)^2} \right) \enspace ,
\end{equation*}
with
\begin{equation*}
\ParVec{\gamma} = \left[ w(\gamma), \, m(\gamma), \, \sigma(\gamma)  \right]^T \enspace .
\end{equation*}

%
%

The required derivatives of the continuous part $f_c^e$ of the CoDiCo-approximation with respect to the density parameters collected in the vector $\vec{p}(x,\gamma)$ given by
\begin{equation*}
\vec{p}(x,\gamma) = \frac{\partial f_c^e(x,\ParVec{\gamma})}{\partial \, \ParVec{\gamma}} =
\begin{bmatrix}
\frac{\partial f_c^e(x,\ParVec{\gamma})}{\partial \, w(\gamma)} \\[2mm]
\frac{\partial f_c^e(x,\ParVec{\gamma})}{\partial \, m(\gamma)} \\[2mm]
\frac{\partial f_c^e(x,\ParVec{\gamma})}{\partial \, \sigma(\gamma)}
\end{bmatrix}
\end{equation*}
or
\begin{equation}
\vec{p}(x,\gamma) =
\begin{pmatrix}
\frac{1}{w(\gamma)} \\[2mm]
\frac{x-m(\gamma)}{\left( \sigma(\gamma) \right)^2} \\[2mm]
\frac{(x-m(\gamma))^2-(\sigma(\gamma))^2}{\left( \sigma(\gamma) \right)^3}
\end{pmatrix}
\, f_c^e(x,\ParVec{\gamma}) \enspace .
\label{Eq_p_vec_scalar}
\end{equation}

%
%

For this specific choice of continuous representation $f_c^e$, the matrix $\Pmat$ is given by
\begin{equation*}
\Pmat=
\begin{pmatrix}
1 & 0 & 0 \\[2mm]
m(\gamma) & w(\gamma) & 0 \\[2mm]
m^2(\gamma) + \sigma^2(\gamma) & 2 \, w(\gamma) \, m(\gamma) & 2 \, w(\gamma) \, \sigma(\gamma)
\end{pmatrix}
\enspace .
\end{equation*}

Solving for $\ParVecDot{\gamma}$ in \Eq{Eq_Pmat_eta_bvec} could be performed based on the matrix $\Pmat$ directly. However, it is possible to give its inverse in closed form, which will be denoted by
\begin{equation*}
\Qmat = \Pmat^{-1} \enspace ,
\end{equation*}
with
\begin{equation*}
\Qmat =
\begin{pmatrix}
1 & 0 & 0 \\[2mm]
-\frac{m(\gamma)}{w(\gamma)} & \frac{1}{w(\gamma)} & 0 \\[2mm]
\frac{m^2(\gamma)-\sigma^2(\gamma)}{2 \, w(\gamma) \, \sigma(\gamma)} & -\frac{m(\gamma)}{w(\gamma) \, s(\gamma)} & \frac{1}{2 \, w(\gamma) \, \sigma(\gamma)}
\end{pmatrix}
\enspace .
\end{equation*}
Hence, we obtain
\begin{equation}
\ParVecDot{\gamma} = \Qmat \, \bvec \enspace ,
\label{Eq_Explicit_ODE}
\end{equation}
with
\begin{equation*}
\ParVecDot{\gamma} = \left[ \dot{w}(\gamma), \, \dot{m}(\gamma), \, \dot{\sigma}(\gamma)  \right]^T \enspace .
\end{equation*}

%
%

The \rhs{} of \Eq{Eq_Explicit_ODE} that we will now denote by $\vec{r}(\gamma)$ can now be written as
\begin{equation}
\begin{split}
\vec{r}(\gamma) & = \Qmat \, \bvec \\
& = \Qmat \, \int_\NewR \vec{m} \cdot \tilde{f}^e(x,\gamma) \cdot \log\left( f_L(x) \right) \, d x \\
& = \int_\NewR \vec{q}(x,\gamma) \cdot \tilde{f}^e(x,\gamma) \cdot \log\left( f_L(x) \right) \, d x
\enspace ,
\end{split}
\label{Eq_ODE_Explicit_r}
\end{equation}
with
\begin{equation*}
\vec{q}(x,\gamma) =
\begin{pmatrix}
1 \\[1mm]
\frac{x-m(\gamma)}{w(\gamma)} \\[3mm]
\frac{(x-m(\gamma))^2-(\sigma(\gamma))^2}{2 \, w(\gamma) \, \sigma(\gamma)}
\end{pmatrix} \enspace .
\end{equation*}

\subsection*{Approximate Expressions for Vector $\bvec$}

%
%

Calculating $\vec{r}(\gamma)$ in \Eq{Eq_ODE_Explicit_r} amounts to calculating certain nonlinear moments of the true posterior density. as closed-form solutions are a rare occasion, we have to be content with approximate integration, which is pursued further here.

%
%

One viable option would be to replace the prior density $f^p(\vec{x})$ by its Dirac Mixture approximation and thus, turn the integration into summation.
%
%
However, this is only efficient for small values of $\gamma$ as long as the progressive Likelihood function is wide and does not modify the prior density too much. For larger $\gamma$, the Likelihood function typically becomes narrower and narrower and would force many Dirac components to zero, which is similar to the degeneration problem in particle filtering. By doing so, only very few Dirac components would really contribute to the integration, especially for larger $\gamma$, which is not desired.

%
%

Thanks to progressive processing, we have an approximate posterior $f_c^e(\vec{x},\ParVec{\gamma})$ available at every $\gamma$. This approximate posterior now allows to integrate only over those portions of the state space that contains the true posterior. For doing so, we
%
%
replace the continuous part $f_c^e(x,\ParVec{\gamma})$ of the CoDiCo-approximation in \Eq{Eq_ODE_Explicit_r} by the coupled discrete density $f^e_d$, i.e., a scalar Dirac Mixture approximation according to
\begin{equation*}
f^e_d(x,\ParVec{\gamma}) = \sum_{i=1}^{L_d} w_i(\ParVec{\gamma}) \; \delta(x - \hat{x}_i(\ParVec{\gamma})) \enspace .
\end{equation*}
As a result, we obtain
\begin{equation}
\vec{r}(\gamma) = \sum_{i=1}^{L_d} w_i \cdot \vec{q}(\hat{x}_i,\gamma) \, 
\frac{\tilde{f}^e(\hat{x}_i,\gamma)}{f_c^e(\vec{x},\ParVec{\gamma})} \cdot \log\left( f_L(\hat{x}_i) \right) \, d x \enspace ,
\label{Eq_bvec_scalar_DMapprox}
\end{equation}
where the explicit dependency of the parameters of the discrete density representation upon the parameters of the continuous part has been omitted.

%
%

It is important to note that by employing Dirac components solely in regions of the state space where the true posterior $\tilde{f}^e(\hat{x}_i,\gamma)$ is represented by its Gaussian approximation $f_c^e(\vec{x},\ParVec{\gamma})$, all the components contribute to the integration. As a result, by far fewer components are required to achieve a good accuracy.

%
%

For the special case of a scalar measurement equation with additive Gaussian noise leading to a scalar Likelihood function 
\begin{equation*}
f_L(x) = \exp\left( -\frac{1}{2} \frac{(y-h(x))^2}{\sigma_v^2} \right) \enspace ,
\end{equation*}
we obtain
\begin{equation*}
\log\left( f_L(\hat{x}_i) \right) = -\frac{1}{2} \frac{(y-h(\hat{x}_i))^2}{\sigma_v^2} \enspace .
\end{equation*}
for $i=1,\ldots\,L_d$.
	

\section{Evaluation} \label{Sec_Eval}

    %
%

For demonstrating the significant increase in performance achieved by the new filter, it is evaluated in comparison with the state-of-the-art filters. As a benchmark example, we consider the discrete-time cubic sensor problem
\begin{equation*}
y = x^3 + v \enspace ,
\end{equation*}
where $v$ is zero-mean Gaussian measurement noise. This is a canonical example, simple to understand, and, compared to some contrived artificial example, allows a good assessment of filter performance.

%
%

For comparison purposes, the proposed new filter is compared to
\begin{itemize}
\item  the class of Linear Regression Kalman Filters (LRKF), specifically to
\begin{itemize}
\item the Unscented Kalman Filter (UKF),
\item  the Gaussian Filter (GF),
\item  the Analytic LRKF,
\end{itemize}
\item and an assumed-Gaussian density filter using moment matching for fitting a Gaussian to the true posterior employing Monte Carlo integration for approximating the desired moments (MC).
\end{itemize}

\begin{figure*}[ht]
\includegraphics[width=0.99\textwidth]{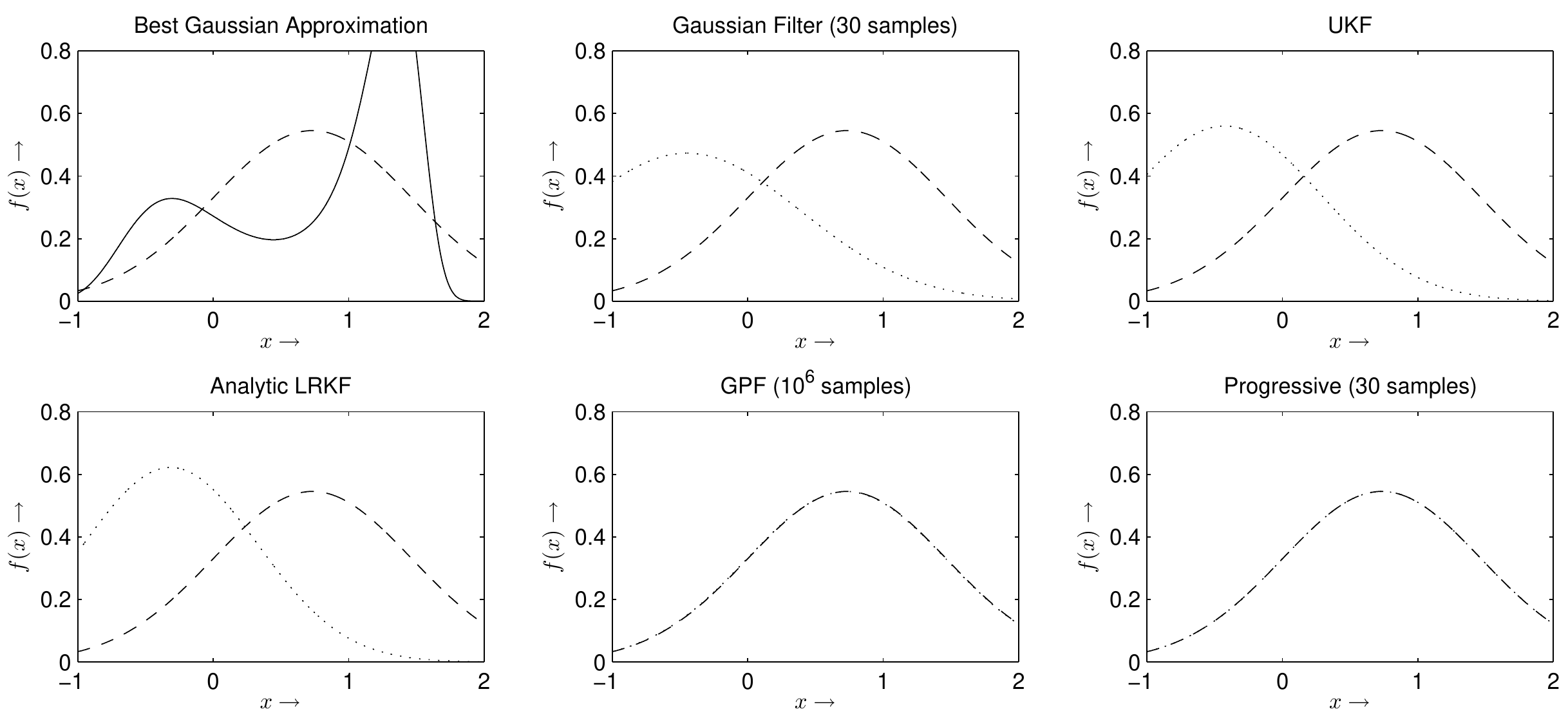}
\caption{Comparison of the results for one filter step for the ground truth, i.e., the best Gaussian approximation computed numerically, compared with the true posterior, the Gaussian Filter (GF), the Unscented Kalman Filter (UKF), the Analytic LRKF, the Gaussian Particle Filter (GPF), and the new progressive Gaussian filter.}
\label{Fig_Comparison_one_step}
\end{figure*}

For a measurement $\hat{y}=3$, a prior Gaussian density with mean $-1$ and variance $1$, and a cubic measurement equation corrupted by zero-mean Gaussian noise with variance $1.2$, the results of one filter step for the different filters are shown in \Fig{Fig_Comparison_one_step}. The true posterior of the Bayesian filter step is calculated on a fine grid with $30.000$ grid points. The ground truth for comparison purposes is the Gaussian density with mean and variance identical to the true posterior. A standard Unscented Kalman Filter (UKF) is employed. The Gaussian Filter (GF) is employed based on an optimal Dirac Mixture approximation of the prior Gaussian density according to 
\cite{IFAC08_Huber}, 
where the filter step is performed by assuming that measurement and state are jointly Gaussian. The Analytic LRKF computes the moments in closed form and additionally assumes that measurement and state are jointly Gaussian. The Gaussian Particle Filter (GPF) computes posterior mean and variance by means of $10^6$ random samples drawn from the prior Gaussian density that are used for evaluating the product of Likelihood and prior density. Subsequently, the resulting weighted samples are used for moment matching. The new progressive Gaussian filter is applied with a discrete density approximation using $30$ Dirac components.

 \Fig{Fig_Comparison_one_step} clearly shows that the LRKFs have difficulties approximating the true posterior. In contrast, the GPF for $10^6$ samples and the new progressive Gaussian filter produce results almost identical to the ground truth.

\begin{figure*}[ht]
\includegraphics[width=0.99\textwidth]{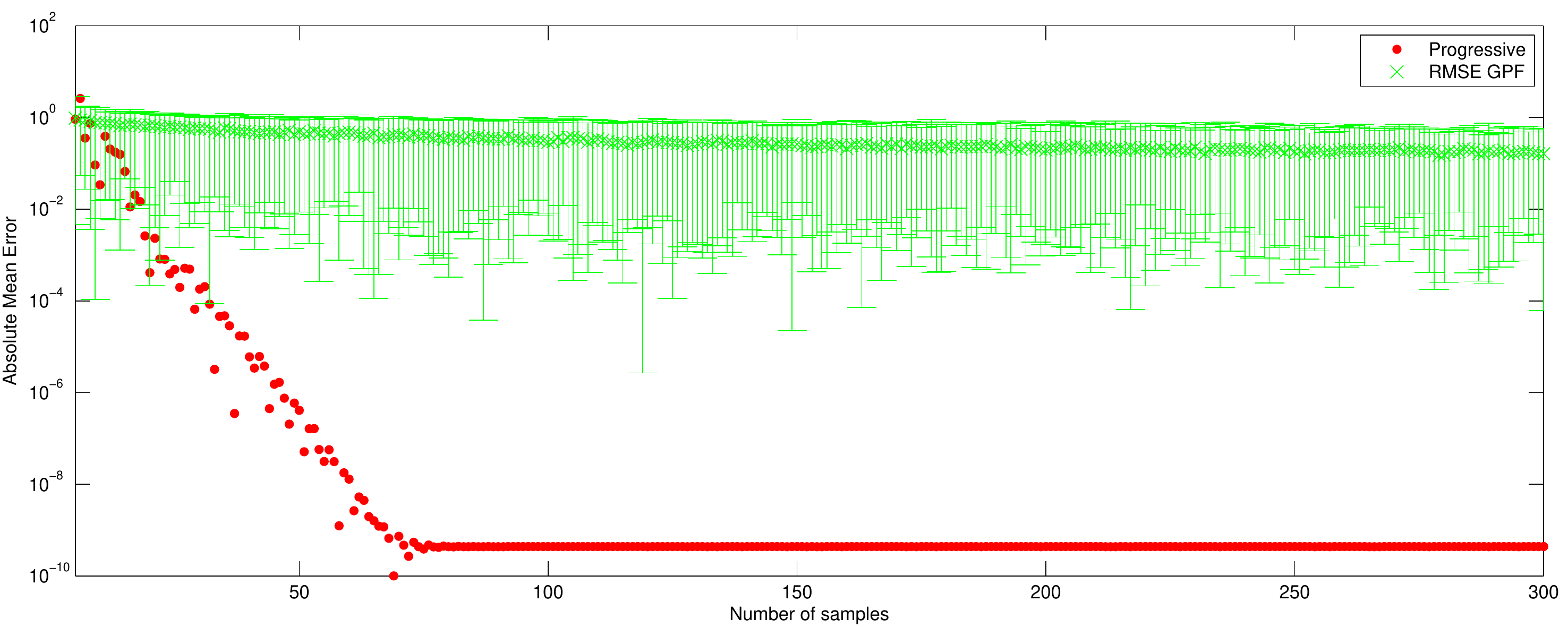}
\caption{Comparison of the estimation quality as a function of the number of samples used between the Gaussian particle filter (GPF) and the proposed progressive Gaussian filter. As the GPF provides non-deterministic results, $100$ Monte-Carlo runs have been performed to produce the RMSE and the min-max error bars.}
\label{Fig_Comparison_num_samples}
\end{figure*}

To illustrate the savings in the number of samples used by the new progressive Gaussian filter, its estimation quality is compared to the Gaussian particle filter in \Fig{Fig_Comparison_num_samples} for the same experimental parameters as above. It is obvious that very good estimation results are already obtained for very few (deterministic) samples. In contrast, the results of the Gaussian particle filter are non-deterministic and the average convergence is much slower.

\begin{figure*}[ht]
\includegraphics[width=0.99\textwidth]{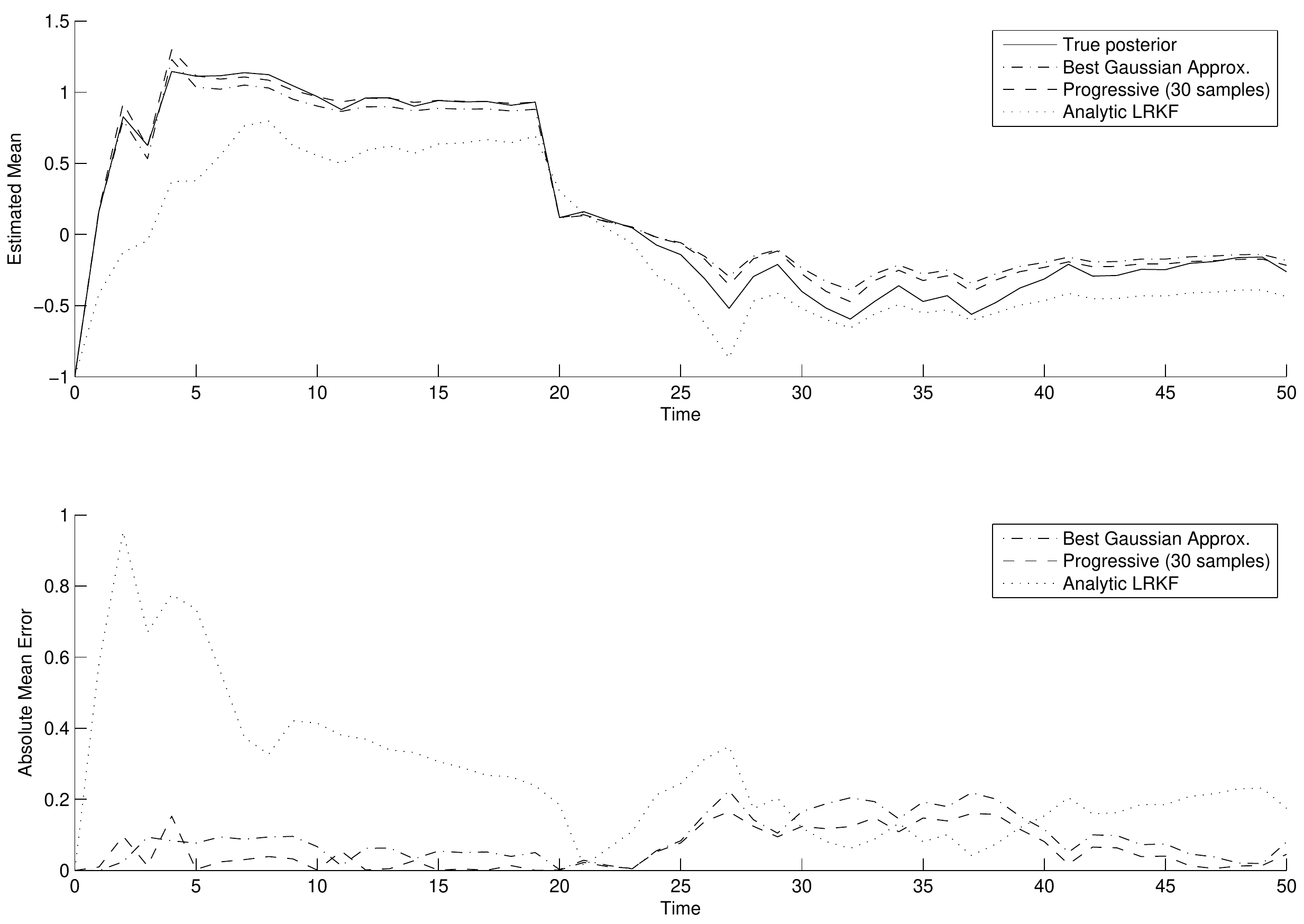}
\caption{Results of recursive filtering over $50$ times steps, where the recursion is started with a prior Gaussian density with mean $-1$ and variance $30$. From time step $1$ to $19$, the true state is $1$, from time step $20$ to $50$, the true state is changed to $0$. Top: Estimated mean sequences. Bottom: Absolute mean error of the best Gaussian approximation, the new progressive filter, and the Analytic LRKF with respect to the mean of the true posterior.}
\label{Fig_Measurement_Sequence}
\end{figure*}

The results of recursive filtering over $50$ times steps are shown in \Fig{Fig_Measurement_Sequence}. The recursion is started with a prior Gaussian density with mean $-1$ and variance $30$. Noise mean is $0$, the noise variance is $1.2$. From time step $1$ to $19$, the true state is $1$. At time step $20$, the true state is changed to $0$ and Gaussian noise with variance of $9$ is added to all estimates.

The top plot in \Fig{Fig_Measurement_Sequence} shows the estimated means of the true posterior, its best Gaussian approximation\footnote{The best Gaussian approximation used as the ground truth at every time step is also recursively calculated on the grid based on the previous best Gaussian approximation. The accumulating error between the true posterior and the ground truth is unavoidable due to the Gaussian assumption.}, the new progressive Gaussian filter, and the Analytic LRKF. The other variants of LRKF have been omitted as the Analytic LRKF provides their lower bound in terms of estimation quality.

The bottom plot in \Fig{Fig_Measurement_Sequence} shows the absolute mean error of the best Gaussian approximation, the new progressive filter, and the Analytic LRKF with respect to the mean of the true posterior.

The two plots in \Fig{Fig_Measurement_Sequence} show that the proposed progressive Gaussian filter provides results very close to the best Gaussian approximation, which itself is rather close to the mean of the true posterior. The analytic LRKFs represented by their highest-quality variant, the Analytic LRKF, show a much larger deviation.

%


\section{Conclusions} \label{Sec_Conclusions}

    %
%
A progressive Gaussian filtering method has been introduced that is based on a \CoDiCo-representation with a Gaussian density as the continuous part and its Dirac Mixture approximation as the discrete part. 
%
%
For the moment calculation, The Dirac Mixture part is employed to evaluate the true posterior density at discrete points only. In contrast to Monte-Carlo methods available for that purpose, the Dirac components all contribute to the integration as progressive processing is exploited to solely place components in the important regions of the state space.
%
%
As a result, the new filter requires significantly less samples, is fast, efficient, and robust, and achieves a high estimation quality compared to state-of-the-art Gaussian filters.
%
%
It can easily be used as a replacement for standard Gaussian filters such as the UKF.
    

\bibliographystyle{IEEEtran}
\bibliography{%
Literature,%
BibTeX-ISAS/ISASPublikationen,%
BibTeX-ISAS/ISASPublikationen_laufend%
}

\end{document}